\documentclass[twocolumn,prl,aps,showpacs,amsmath]{revtex4}
\usepackage{epsfig,amssymb,amsfonts}

\def\opone{\leavevmode\hbox{\small1\kern-3.8pt\normalsize1}}

\begin{document}

\title{Thermal entanglement in a two-spin-qutrit system under a nonuniform external
magnetic field}
\author{Guo-Feng Zhang\footnote{Corresponding author.}\footnote{Email:
gf1978zhang2001@yahoo.com}}\affiliation{State Key Laboratory for
Superlattices and Microstructures, Institute of
Semiconductors,\\Chinese Academy of Sciences, P. O. Box 912,
Beijing 100083, People's Republic of China}
\author{Shu-Shen Li}\affiliation{ CCAST (World Lab.), P.O. Box 8730,
Beijing 100080, and State Key Laboratory for Superlattices and
Microstructures, Institute of Semiconductors, Chinese Academy of
Sciences, P.O. Box 912, Beijing 100083, China}

\begin{abstract}
The thermal entanglement in a two-spin-qutrit system with two
spins coupled by exchange interaction under a magnetic field in an
arbitrary direction is investigated. Negativity, the measurement
of entanglement, is calculated. We find that for any temperature
the evolvement of negativity is symmetric with respect to magnetic
field. The behavior of negativity is presented for four different
cases. The results show that for different temperature, different
magnetic field give maximum entanglement. Both the parallel and
antiparallel magnetic field cases are investigated qualitatively
(not quantitatively) in detail, we find that the entanglement may
be enhanced under an antiparallel magnetic field.
\end{abstract}

 \maketitle

\section{I. introduction}

It is well known that quantum entanglement \cite{aei,esc,jsb}
plays a fundamental role in almost all efficient protocols of
quantum computation (QC) and quantum information processing
\cite{chb,man}. In one proposal \cite{bek} for physical
implementation of qubits, a well localized nuclear spin coupled
with an electron of a donor atom in silicon plays the role of a
qubit which can be individually initialized, manipulated and read
out by extremely sensitive devices. In another
proposal\cite{dlo,div,gbu,aim}, the spin of an electron in a
quantum dot plays the role of a qubit. Long decoherence time and
scalability to more than 100 qubits are two of the important
virtues of both the schemes. In both schemes the effective
interaction between the two qubits is governed by an isotropic
Heisenberg Hamiltonian with Zeeman coupling of the individual
spins, namely
\begin{equation}
\emph{H}=JS_{1}\cdot S_{2}+\gamma (S_{1z}+S_{2z}).
\end{equation}
At extremely low temperatures such a qubit system may be assumed
to be in its ground state. However a real physical system is
always at a finite temperature and hence in a mixture of
disentangled and entangled states depending on the temperature.
Therefore one is naturally led to consider the thermal
entanglement of such physical systems. The thermal entanglement
has been extensively studied for various systems including
isotropic \cite{mca,kmo,xwa,gfz} and anisotropic \cite{xwan}
Heisenberg chains, Ising model in an arbitrarily directed magnetic
field \cite{dgu}, and cavity-QED \cite{sma} since the seminal
works by Arnesen et al.\cite{mca} and Nielsen \cite{mani}. Based
on the method developed in the context of quantum information, the
relaxation of a quantum system towards the thermal equilibrium is
investigated \cite{vsc} and provides us an alternative mechanism
to model the spin systems of the spin-$\frac 12$ case for the
approaching of the thermal entangled
states\cite{mca,kmo,xwa,gfz,xwan}. It should be noted that only
the uniform field case is carefully studied in the above-mentioned
papers. The nonuniform case is rarely taken into account. But in
any solid state construction of qubits, there is always the
possibility of inhomogeneous Zeeman coupling \cite{xhu,xhur}.
Moreover for perform quantum computing, it is necessary to control
the magnetic field at each spin separately\cite{ymakhlin}. So in
the theoretical analysis, the nonuniform external magnetic field
should be included in the model Hamiltonian. Recently, Sun
\cite{ysu} and M.Asoudeh \cite{mas} investigate the thermal
entanglement in the two-qubit spin model with a nonuniform
magnetic field. But only the spin-$\frac 12$ case is carefully
studied in the above papers. Zhang et al.\cite{gzh} only consider
the uniform magnetic field for spin-$1$ case. In this paper, we
will investigate the thermal entanglement in the two-spin-$1$
system with a magnetic field in an arbitrary direction. Thus we
may better understand and make use of entanglement in quantum
information processing through changing the environment.

Our paper is arranged as follows: first we will give the
definition of negativity, the measurement of entanglement. After
giving the model Hamiltonian and the solutions, we will present
our calculation results by several figures. Finally, the
discussion and conclusion remarks will be given.

\section{II. The definition of negativity}

We first introduce the concept of negativity, which will be used
as the entanglement measure. The Peres-Horodecki
criterion\cite{ape} gives a qualitative way for judging if the
state is entangled. The quantitative version of the criterion was
developed by Vidal and Werner\cite{gvi}. They presented a measure
of entanglement called negativity that can be computed
efficiently, and the negativity does not increase under local
manipulations of the system. The negativity of a state $\rho $ is
defined as
\begin{equation}
N(\rho )=\sum_i\left| \mu _i\right|,
\end{equation}
where $\mu _i$ is the negative eigenvalue of $\rho ^{T_1}$, and
$T_1$ denotes the partial transpose with respect to the first
system. The negativity $N$ is related to the trace norm of $\rho
^{T_1}$ via\cite{gvi}
\begin{equation}
N(\rho )=\frac{\left| \left| \rho ^{T_1}\right| \right| _1-1}2.
\end{equation}
where the trace norm of $\rho ^{T_1}$ is equal to the sum of the
absolute values of the eigenvalues of $\rho ^{T_1}$. If $N>0$,
then the two-spin state is entangled.

The state of a system at thermal equilibrium can be described by
the density operator $\rho (T)=\exp (-\beta H)/Z$, where
$Z=Tr[\exp (-\beta H)]$ is the partition function and $\beta
=1/k_BT$ ($k_B$ is Boltzmann's constant being set to be unit
$k_B=1$ hereafter for the sake of simplicity and $T$ is the
temperature). The entanglement in the thermal state is called
thermal entanglement.
\section{III.The model Hamiltonian and the solutions}
The development of laser cooling and trapping provides us more
ways to control the atoms in traps. Indeed, we can manipulate the
atom-atom coupling constants and the atom number in each lattice
well with a very good accuracy. Our system consists of two wells
in the optical lattice with one spin-1 atom in each well. The
lattice may be formed by three orthogonal laser beam, and we may
use an effective Hamiltonian of the Bose-Hubbard form\cite{dja}to
describe the system. The atoms in the Mott regime make sure that
each well contains only one atom. For finite but small hopping
term $t$, we can expand the Hamiltonian into powers of $t$ and
get\cite{sky},
\begin{equation}
H=\epsilon +J(S_1\cdot S_2)+K(S_1\cdot S_2)^2,
\end{equation}
where $J=-\frac{2t^2}{U_2}$,
$K=-\frac{2t^2}{3U_2}-\frac{4t^2}{U_0}$ with $t$ the hopping
matrix elements, and $\epsilon =J-K$. $U_s$($s=0,2$) represents
the Hubbard repulsion potential with total spin $s$, a potential $U_s$ with $%
s=1$ is not allowed due to the identity of the bosons with one
orbital state per well, Since term $\epsilon $ contains no
interaction, we can ignore it in the following discussions and it
would not change the thermal entanglement. For simplification,
$J\gg K$ is assumed and the nonlinear couplings is ignored. So the
Hamiltonian Eq.(4) becomes
\begin{equation}
H=J(S_1\cdot S_2),
\end{equation}
with an nonuniform external magnetic field in an arbitrary
direction, our system is described by
\begin{equation}
H=J(S_{1x}S_{2x}+S_{1y}S_{2y})+B\cos[\theta]S_{1z}+B\sin[\theta]S_{2z}.
\end{equation}
in which the neglected exchange coupling term along the $z$-axes
is assumed
to be much smaller than the coupling in the x-y plane. Where $S_\alpha $ ($%
\alpha =x,y,z$) are the spin operator, $J$ is the strength of
Heisenberg interaction and the magnetic field is assumed to be
along the $z$-axes. When the total spin for each site $S_j=1$
$(j=1,2)$, its components take the form,
\begin{eqnarray}
S_{jx}&=&\frac 1{\sqrt{2}}\left(
\begin{array}{lll}
0 & 1 & 0 \\
1 & 0 & 1 \\
0 & 1 & 0
\end{array}
\right),\nonumber
\\S_{jy}&=&\frac 1{\sqrt{2}}\left(
\begin{array}{lll}
0 & -i & 0 \\
i & 0 & -i \\
0 & i & 0
\end{array}
\right), \nonumber
\\S_{jz}&=&\left(
\begin{array}{lll}
1 & 0 & 0 \\
0 & 0 & 0 \\
0 & 0 & -1
\end{array}
\right).
\end{eqnarray}
In the following calculation we will select $J$ as the energy unit and set $%
J=1$.

To evaluate the thermal entanglement we first of all find the
eigenvalues and the corresponding eigenstates of the Hamiltonian
Eq.(6). which are seen to be
\begin{eqnarray}
H|\psi_{1}\rangle&=&0,
 H|\psi_{2}\rangle=-B_{+}|\psi_{2}\rangle,\nonumber
\\H|\psi_{3}\rangle&=&B_{+}|\psi_{3}\rangle,
 H|\psi_{4}^{\pm}\rangle=-m_{\mp}|\psi_{4}^{\pm}\rangle,\nonumber
\\H|\psi_{5}^{\pm}\rangle&=&m_{\pm}|\psi_{5}^{\pm}\rangle,
H|\psi_{6}^{\pm}\rangle=\pm\xi|\psi_{6}^{\pm}\rangle.
\end{eqnarray}
where $B_{\pm}=B\cos[\theta]\pm B\sin[\theta]$,
$\xi=\sqrt{2+B_{-}^{2}}$, $\zeta=\sqrt{4+B_{-}^{2}}$,
$m_{\pm}=\frac{1}{2}(B_{+}\pm\zeta)$.

And the corresponding eigenstates are explicitly given by
\begin{eqnarray}
|\psi_{1}\rangle&=&\frac{1}{\xi}(|-1,1\rangle+B_{-}|0,0\rangle-|1,-1\rangle),\nonumber
\\|\psi_{2}\rangle&=&|-1,-1\rangle,\nonumber
\\ |\psi_{3}\rangle&=&|1,1\rangle,\nonumber
\\|\psi_{4}^{\pm}\rangle&=&\frac{1}{\sqrt{1+S_{\pm}^{2}}}(|-1,0\rangle+S_{\pm}|0,-1\rangle),\nonumber
\\|\psi_{5}^{\pm}\rangle&=&\frac{1}{\sqrt{1+S_{\pm}^{2}}}(|0,1\rangle+S_{\pm}|1,0\rangle),\nonumber
\end{eqnarray}
\begin{widetext}
\begin{equation}
|\psi_{6}^{\pm}\rangle=\frac{1}{\sqrt{1+R_{\pm}^{2}+(1\pm
B_{-}R_{\pm})^{2}}}[|-1,1\rangle\pm R_{\pm}|0,0\rangle+(1\pm
B_{-}R_{\pm})|1,-1\rangle].
\end{equation}
\end{widetext}
where $R_{\pm}=\pm B_{-}+\xi$ and
$S_{\pm}=\frac{1}{2}(B_{-}\pm\zeta)$. Here $\left|
x,y\right\rangle $ ($x=1,0,-1$ and $y=1,0,-1$) are the eigenstates
of $S_{1z}S_{2z}$. The density operator $\rho$ can be expressed in
terms of the eigenstates and the corresponding eigenvalues as
\begin{equation}
\rho =\frac 1Z\sum\exp [-\beta E_l]\left| \Psi _l\right\rangle
\left\langle \Psi _l\right|,
\end{equation}
where $E_l$  is the eigenvalue of the corresponding eigenstates
and the partition function is $Z=1+2\cosh [\beta \xi ]+4\cosh
[\frac{1}{2}\beta\zeta ]\cosh [\frac{\beta B_{+}}{2} ]+2\cosh
[\beta B_{+}]$.
\begin{figure*}
\begin{center}
\epsfig{figure=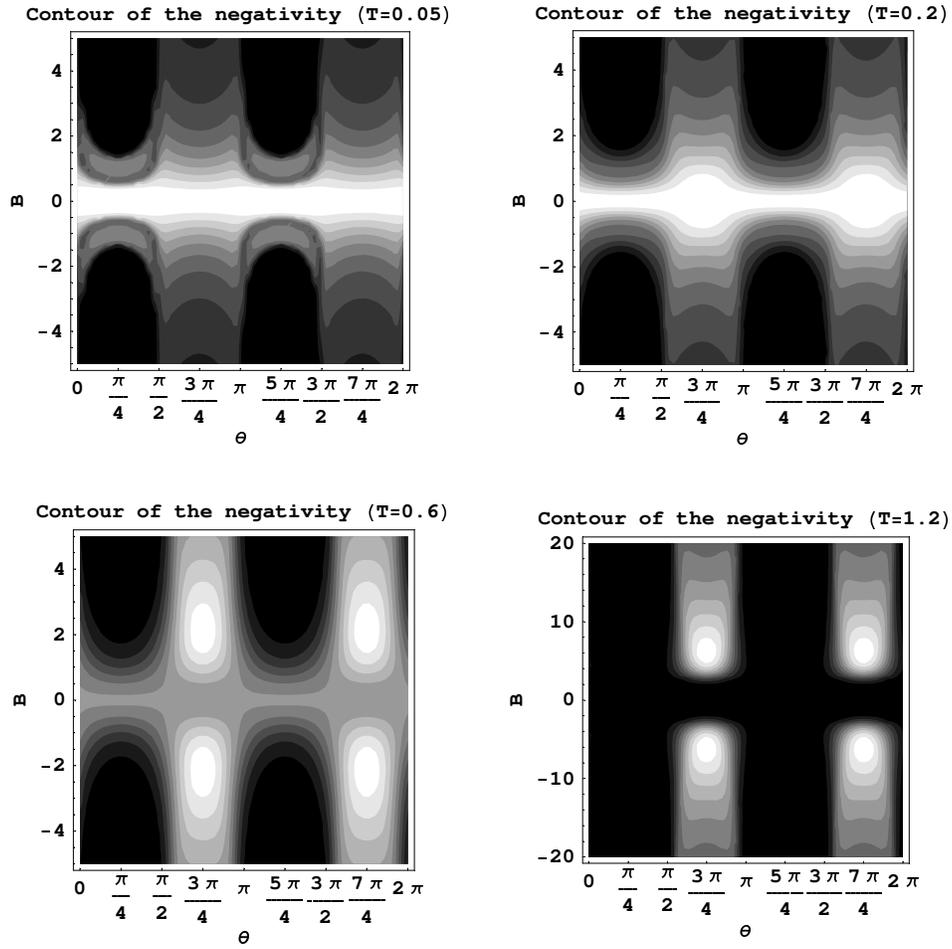,width=0.70\textwidth}
\end{center}
\caption{The contour of negativity for different temperature vs.
$B$ and $\theta$. The brighter place means the higher negativity.}
\label{FIG1}
\end{figure*}
\begin{figure*}
\begin{center}
\epsfig{figure=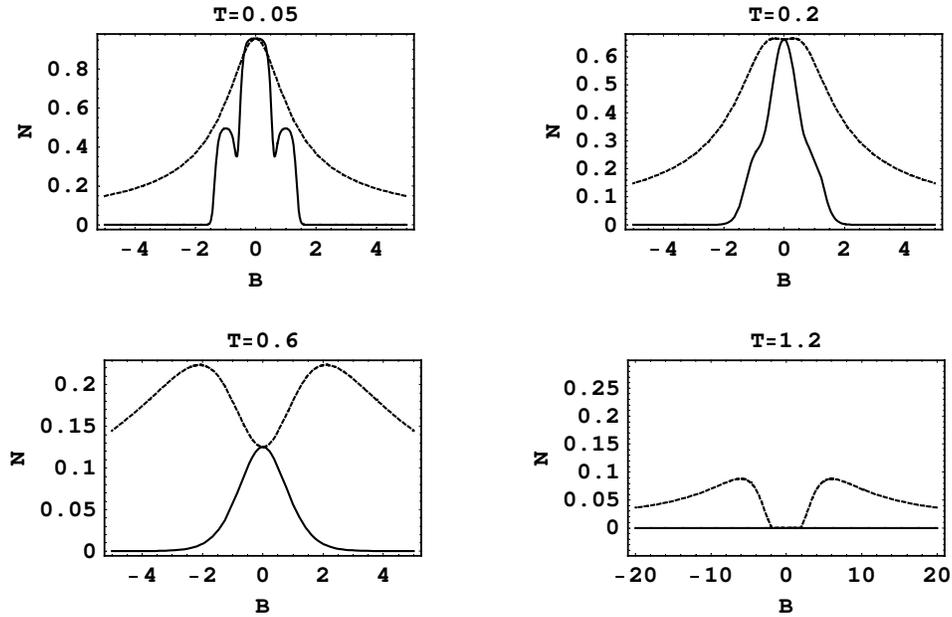,width=0.70\textwidth}
\end{center}
\caption{The negativity versus $B$ for different temperature.
solid curve for $\theta=\frac{\pi}{4}$ and dotted curve for
$\theta=\frac{3\pi}{4}$. } \label{FIG2}
\end{figure*}

For our purpose to evaluate the negativity in what following we
need to have a partially transposed density matrix $\rho ^{T_1}$
of original density matrix $\rho $ with respect to the eigenbase
of any one spin particle ( say particle $1$) which is found in the
basis $\left| x,y\right\rangle $ ($x=1,0,-1$ and $y=1,0,-1$) as
\begin{widetext}
\begin{equation}
\rho ^{T_1}=\frac{1}{Z}\left(%
\begin{array}{ccccccccc}
  e^{-\beta B_{+}} & 0 & 0 & 0 & q_{+} & 0 & 0 & 0 &   \frac{-1+\cosh[\beta \xi]}{\xi^{2}} \\
  0 & Me^{-\frac{\beta B_{+}}{2}} & 0 & 0 & 0 & u_{-} & 0 & 0 & 0 \\
  0 & 0 & W_{-} & 0 & 0 & 0 & 0 & 0 & 0 \\
  0 & 0 & 0 & Q_{-} & 0 & 0 & 0 & u_{+} & 0 \\
  q_{+} & 0 & 0 & 0 & 1+2(\frac{-1+\cosh[\beta \xi]}{\xi^{2}}) & 0 & 0 & 0 & q_{-} \\
  0 & u_{-} & 0 & 0 & 0 & Me^{\frac{\beta B_{+}}{2}} & 0 & 0 & 0 \\
  0 & 0 & 0 & 0 & 0 & 0 & W_{+} & 0 & 0 \\
  0 & 0 & 0 & u_{+} & 0 & 0 & 0 & Q_{+} & 0 \\
  \frac{-1+\cosh[\beta \xi]}{\xi^{2}} & 0 & 0 & 0 & q_{-} & 0 & 0 & 0 & e^{\beta B_{+}} \\
\end{array}%
\right)
\end{equation}
\end{widetext}
 where $M=\cosh
[\frac{1}{2}\beta\zeta]-\frac{\sinh[\frac{1}{2}\beta\zeta]B_{-}}{\zeta}$
, $q_{\pm }=-\frac{1}{\zeta}[e^{-\frac{1}{2}}\beta(\zeta\pm
B_{+})(-1+e^{\beta \zeta})]$, $u_{\pm}=\frac{\pm
B_{-}(1-\cosh[\beta\xi])-\xi\sinh[\beta\xi]}{\xi^{2}}$, $W_{\pm
}=\frac{1}{\xi^{2}}(1+\cosh[\beta\xi](1+B_{-}^{2})\pm \xi
B_{-}\sinh[\beta\xi])$, $Q_{\pm
}=\frac{1}{2}e^{\frac{1}{2}\beta(\zeta\pm
B_{+})}(1+\frac{B_{-}}{\zeta})+\frac{2e^{\pm\frac{1}{2}\beta (\mp
\zeta+B_{+})}}{4+B_{-}(B_{-}+\zeta)}$.

We perform the numerical diagonalization of the density matrix and
the numerical results of the entanglement measure $N$ is
presented. In Fig.1, we give the contour of negativity for
different temperature with respect to $B$ and $\theta$. From
Fig.1, we can see that the evolvement of negativity is symmetric
with respect to magnetic field. The maximum negativity arrives at
the point $B=0$ when $T=0.05$ and $T=0.2$. With the increasing
temperature, the area of $B$ at which the system can reach maximum
negativity becomes narrower and even arrives zero for a higher
temperature (for example $T=0.6$ and $T=1.2$). When the
temperature is low, only one peak appears. For $T=0.6$ and
$T=1.2$, the double peak structure takes place. We can also find
that the evolvement of negativity is periodic with respect to the
polar angle $\theta$ and the double peak structure takes place at
$\theta=(n+\frac{3}{4})\pi$, ($n=0,1,...$). From these figures, we
know that the negativity gets smaller with the increasing magnetic
field amplitude. For a higher temperature (say $T=1.2$), we can
see that $N$ arrives zero near $B=0$. But $N$ increases with the
value of $B$ to a peak value for $\theta=(n+\frac{3}{4})\pi$
cases, after this peak, $N$ will decrease monotonously.

In order to see clearly the change of the negativity, we give the
results for $\theta=\frac{\pi}{4}$ and  $\theta=\frac{3\pi}{4}$.
This corresponds to the parallel and antiparallel magnetic field
case. We give our calculation results in Fig.2 (solid line for
$\theta=\frac{\pi}{4}$, dotted line for $\theta=\frac{3\pi}{4}$),
in which the negativity is plotted in the whole parameter space at
a given temperature, and four typical cases are shown. As the
temperature is low ($T=0.05$), we may find that there are two
features. First, there are three sharp peaks (different from the
spin-$\frac 12$ case for which only one peak appears, the results
for spin-$\frac{1}{2}$ case can be seen from Ref [23]) and the
center of the middle one locates at $B=0$, where the negativity is
about $1$. As we increase the external field $B$, $N$ rapidly
decays. When $T=0.6$, we can see that the three peaks evolve into
one. That is to say, the left and the right peak disappear, the
middle peak gets shorter as we increase the temperature. As the
temperature is further increased, for example $T=1.2$, the
entanglement is entirely destroyed.

Compared with the parallel magnetic field case
($\theta=\frac{\pi}{4}$), when at the low temperature, for
antiparallel magnetic field ($\theta=\frac{3\pi}{4}$) case, there
is no three-peaks structure emerges and only a peak that
monotonously decreases with the value of $B$. But the negativity
decreases more slowly than in the parallel case which means that
in the strong field region, the parallel field and the
antiparallel field demonstrate obviously different effects on the
entanglement. We can find that in all parameter space, the
negativity of the magnetic field with antiparallel direction is
much larger than that of parallel field. These results can be seen
in these figures. As the temperature increase ($T=0.6$), the
feature of $N$ will be changed. The primary peak at $T=0.05$
splits into two peaks. For the parallel field case, the maximum
$N$ appears at $B=0$, but for antiparallel field case, at the
point $N$ is a minimum point. The results is the same with the one
which can be found in Ref [23]. For $T=0.6$, if we apply an
antiparallel field, $N$ will be enhanced more than two times. This
again illuminates the fact that the well-chosen external field can
partially weaken the destructive effect of thermal fluctuation and
enhance the entanglement. In other words, for a certain
temperature, a well-chosen external field is helpful for
entanglement. At a higher temperature ($ T=1.2$), we can see that
$N$ arrives zero at $B=0$ for the both case. But $N$ increases
with the value of $B$ to a peak value for the antiparallel case,
after this peak, $N$ will decrease monotonously.
\section{IV. conclusions}
We investigated qualitatively (not quantitatively) the effects of
a magnetic field in an arbitrary direction on the thermal
entanglement in the two-spin-$1$ system in terms of the measure of
entanglement called ``negativity". We give results for different
temperatures. We find that the temperature and the magnetic field
can affect the feature of the thermal entanglement significantly.
At a certain temperature, the antiparallel magnetic field is
helpful for entanglement. In other words, the entanglement may be
enhanced under an antiparallel magnetic field.

\end{document}